\documentclass[slac_one]{revtex4}
\usepackage{graphicx}
\usepackage{subfigure}
\usepackage{url}
\usepackage{hyperref}
\usepackage{fancyhdr}
\begin{document}

\title{$\mathrm{W}$ and $\mathrm{Z}$ Production in $pp$
  Collisions at $7\,\mathrm{TeV}$ with ATLAS} 

\author{S. Prasad}
\affiliation{Harvard University, Cambridge, MA 02138, USA}

\begin{abstract}
Measurements of $\mathrm{W}$ and
$\mathrm{Z}$ cross-sections in $pp$ collisions at $E_{\mathrm{CM}}
= 7\,\mathrm{TeV}$ at the Large Hadron Collider are reported from the ATLAS experiment. From an observation of $118$ leptonic $\mathrm{W}$ candidates, the inclusive
$\mathrm{W}$ cross-section times leptonic branching fraction is measured
as $\left[9.3\pm0.9\mathrm{(stat)}\pm0.6\mathrm{(syst)}\pm1.0\mathrm{(lumi)}\right]\mathrm{nb}$. The
result for the $Z$ boson is
$\left[0.83\pm0.07\mathrm{(stat)}\pm0.06\mathrm{(syst)}\pm0.09\mathrm{(lumi)}\right]\mathrm{nb}$. These
results agree with theoretical expectations from NNLO QCD.
\end{abstract}

\maketitle

\thispagestyle{fancy}

\section{INTRODUCTION, DATA SAMPLE AND OBJECT DEFINITIONS}
Study of $\mathrm{W}$ and $\mathrm{Z}$ bosons at the
turn-on of the Large Hadron Collider provides a high-statistics probe of the ATLAS
\cite{atlas} detector, in terms of characterizing tracking, lepton
identification, lepton and missing energy scale and resolution, and
trigger performance. Theoretically, measurements at the new energy of
$7\,\mathrm{TeV}$ probe the proton at low-$x$, constraining parton density functions. The
$\mathrm{W}$ and $\mathrm{Z}$ cross-sections are known at NNLO, giving
a test of perturbative calculations. Also, these channels are important
backgrounds for beyond SM processes. The data for these analyses come from the first LHC runs at
$7\,\mathrm{TeV}$, between March and July 2010. The datasets for the $W$ and $Z$ boson analyses
are approximately $17\,\mathrm{nb}^{-1}$ and $225\,\mathrm{nb}^{-1}$
respectively, the luminosity being to known to $11\%$. 

The analyses operate upon events from a collision bunch and both detector and
machine status are used to select run periods for analysis. The data
are compared to PYTHIA \cite{pythia} samples processed with the ATLAS
detector simulation software based on GEANT4 \cite{geant4} and fully reconstructed. The trigger for the muon analysis uses hit patterns in dedicated
chambers($|\eta|<2.4$). The trigger
efficiency is estimated at $88\%$ (fully dominated by detector
acceptance) using orthogonal jet-based
triggers. The electron data are triggered in calorimetric towers that
identify $e/\gamma$ activity in a window of $|\eta| <
2.5$. A threshold of five trigger counts ($5\,\mathrm{GeV}$)
is used. The efficiency of this trigger is better than $99.9\%$,
with negligible uncertainty. Both $e$ and $\mu$ analyses require a reconstructed vertex ($\ge$ three tracks) consistent with the
beam spot. In addition, to reject cosmic rays and
beam halo events, the muon analyses require that the primary vertex be
found within $15\,\mathrm{cm}$ of the nominal position.

Electron measurements come from the energy of calorimeter clusters
found in using a sliding-window approach. For electron identification, 
calorimetric shower shapes, hadronic leakage, track-cluster matching
and impact parameter criteria are used. For the $W$ analysis, purity is enhanced by requiring $E/p$
compatibility between track and cluster, high-threshold transition radiation hits,
and criteria to reject conversions. The analysis
uses electrons with $p_{T} > 20\,\mathrm{GeV}$ in the region
$|\eta|<2.47$. The barrel-endcap transition region $|\eta| \in \left[1.37,
  1.52\right)$ is excluded. Muon reconstruction combines tracking
information in the Inner Detector (ID) the Muon Spectrometer (MS). Decays in flight are rejected
with minimum $p_{\mathrm{T}}$ requirements on the MS track, and by imposing $p_{\mathrm{T}}$ compatibility
between ID and MS track reconstructions. The cosmic ray contamination is reduced
by requiring that the muon come from within 1 cm of a reconstructed vertex. Muons with $p_{T} > 20\,\mathrm{GeV}$ are used, with
$|\eta|<2.4$. To reduce QCD background, the summed $p_\mathrm{T}$ of tracks in a cone ($\Delta R = 0.4$) around the muon
must be $<0.2$ of muon $p_\mathrm{T}$. Missing $E_{\mathrm{T}}$ is found using EM-scale calorimeter energy deposited in
topological clusters, corrected for hadronic response, dead material
and out-of-cluster losses. For muons, this is supplemented with the
momentum, the muon's calorimeter deposition not being added to the
calorimeter term. A requirement of $25\,\mathrm{GeV}$ is imposed for
the $\mathrm{W}$ analysis.

\section{EVENT SELECTION AND BACKGROUNDS}
\begin{figure}[h!]
\centering
\subfigure[$\mathrm{W}\to e\nu$] {
\label{fig:mtwe}
\includegraphics[width=.4\textwidth]{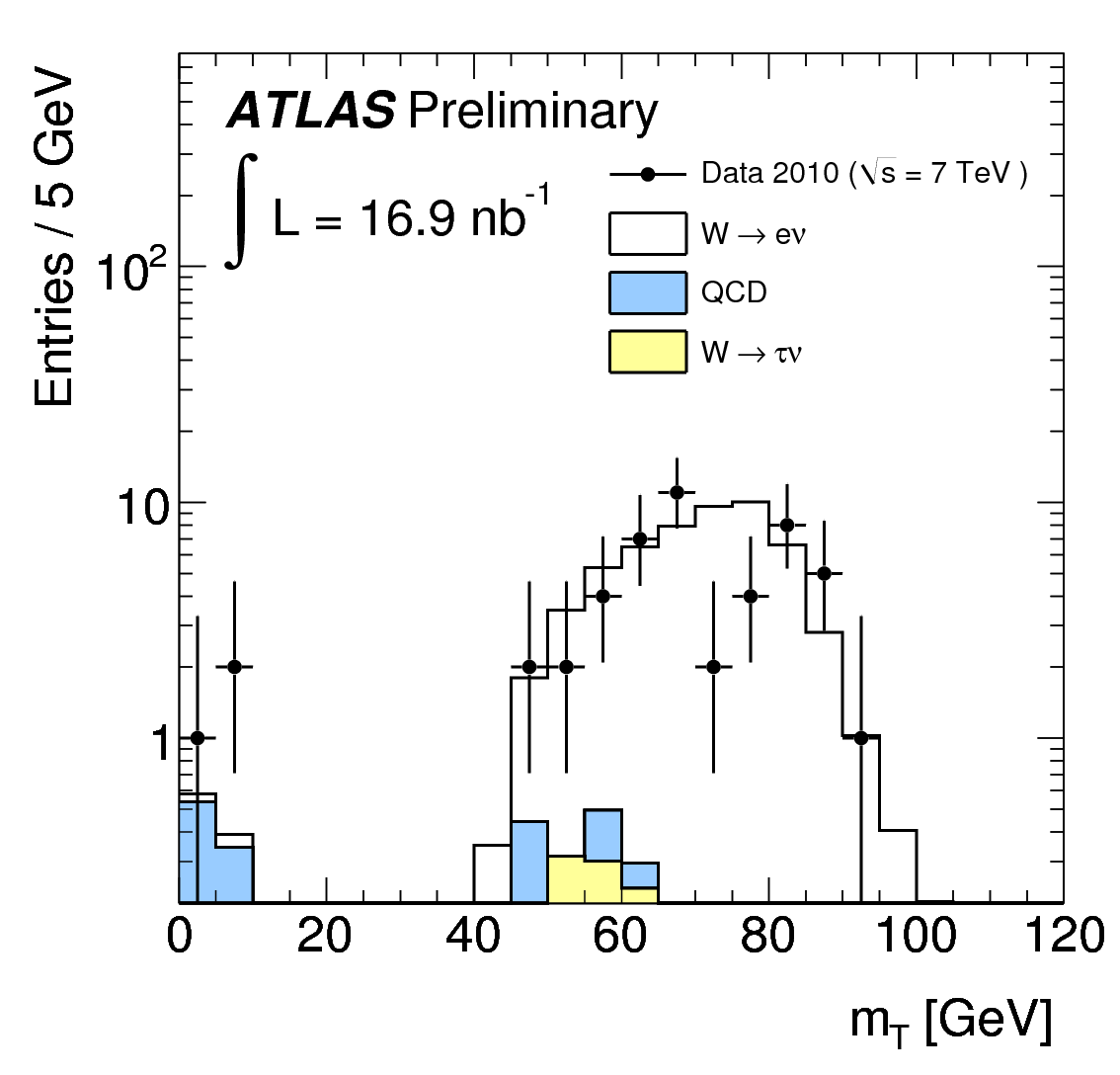}
}
\subfigure[$\mathrm{W}\to \mu\nu$]{
\label{fig:mtwmu}
\includegraphics[width=.4\textwidth]{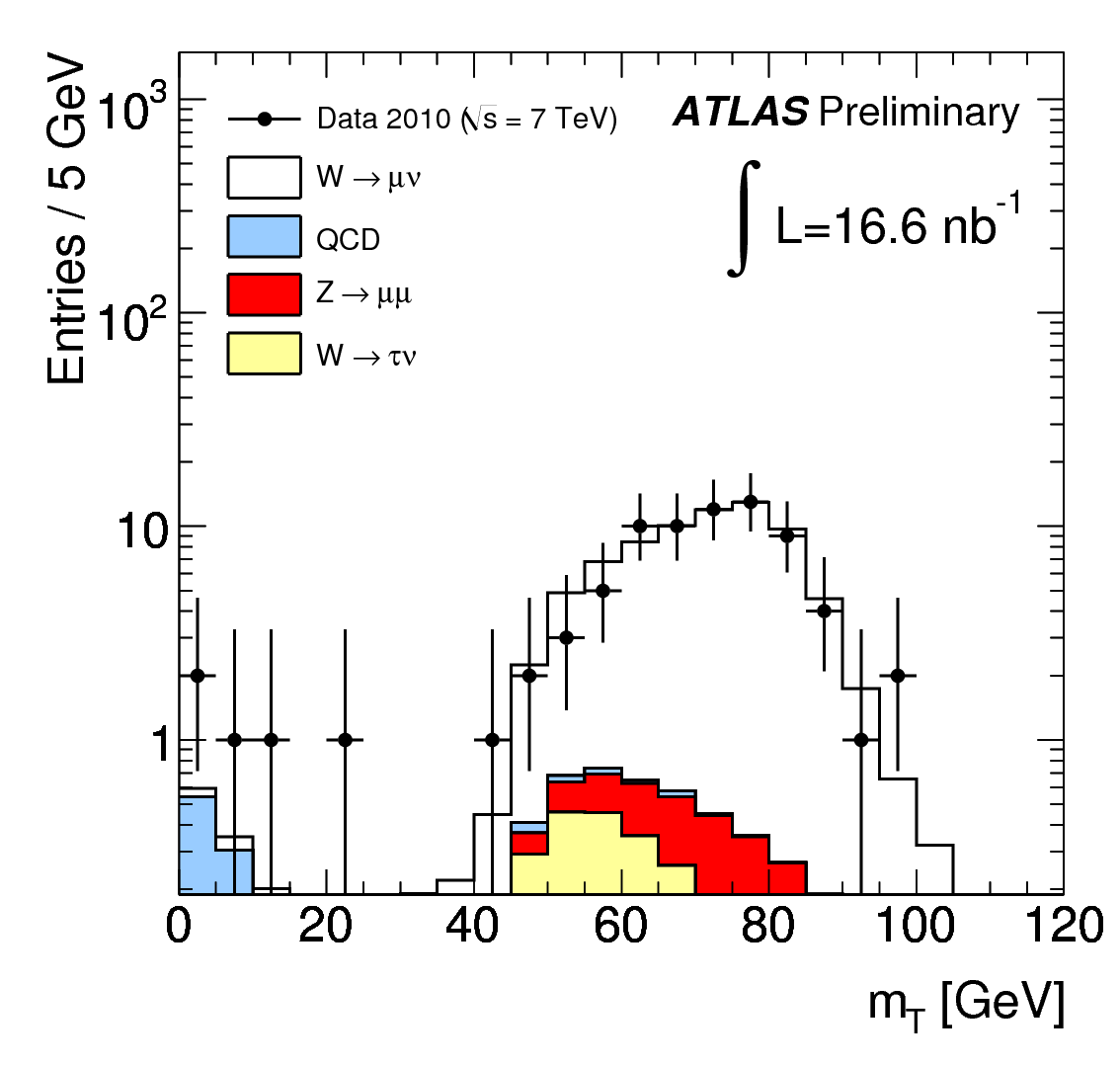}
} \\
\caption{Transverse mass $m_{\mathrm{T}}$ of the (a) $e-E_{T}^{\mathrm{Miss}}$
  and (b) $\mu-E_{T}^{\mathrm{Miss}}$ systems, where $p_\mathrm{T}(l)>20\,\mathrm{GeV}$ and $E_{T}^{\mathrm{Miss}}>25\,\mathrm{GeV}$. \label{fig:mtw}}
\end{figure}

\begin{figure}[h!]
\centering
\subfigure[$\mathrm{Z}\to e^+ e^-$] {
\label{fig:mee}
\includegraphics[width=.4\textwidth]{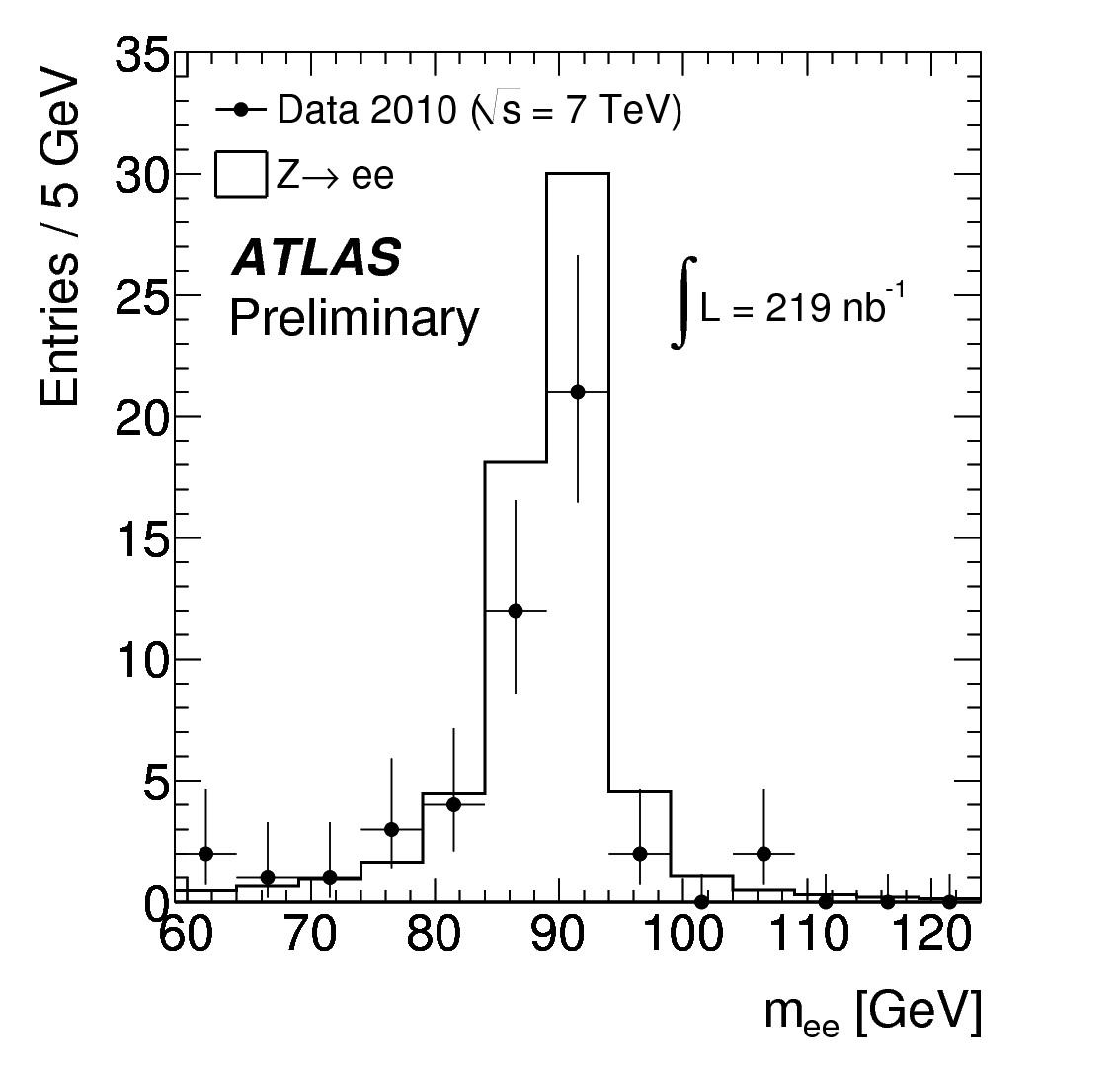}
}
\subfigure[$\mathrm{Z}\to \mu^+\mu^-$]{
\label{fig:mmumu}
\includegraphics[width=.4\textwidth]{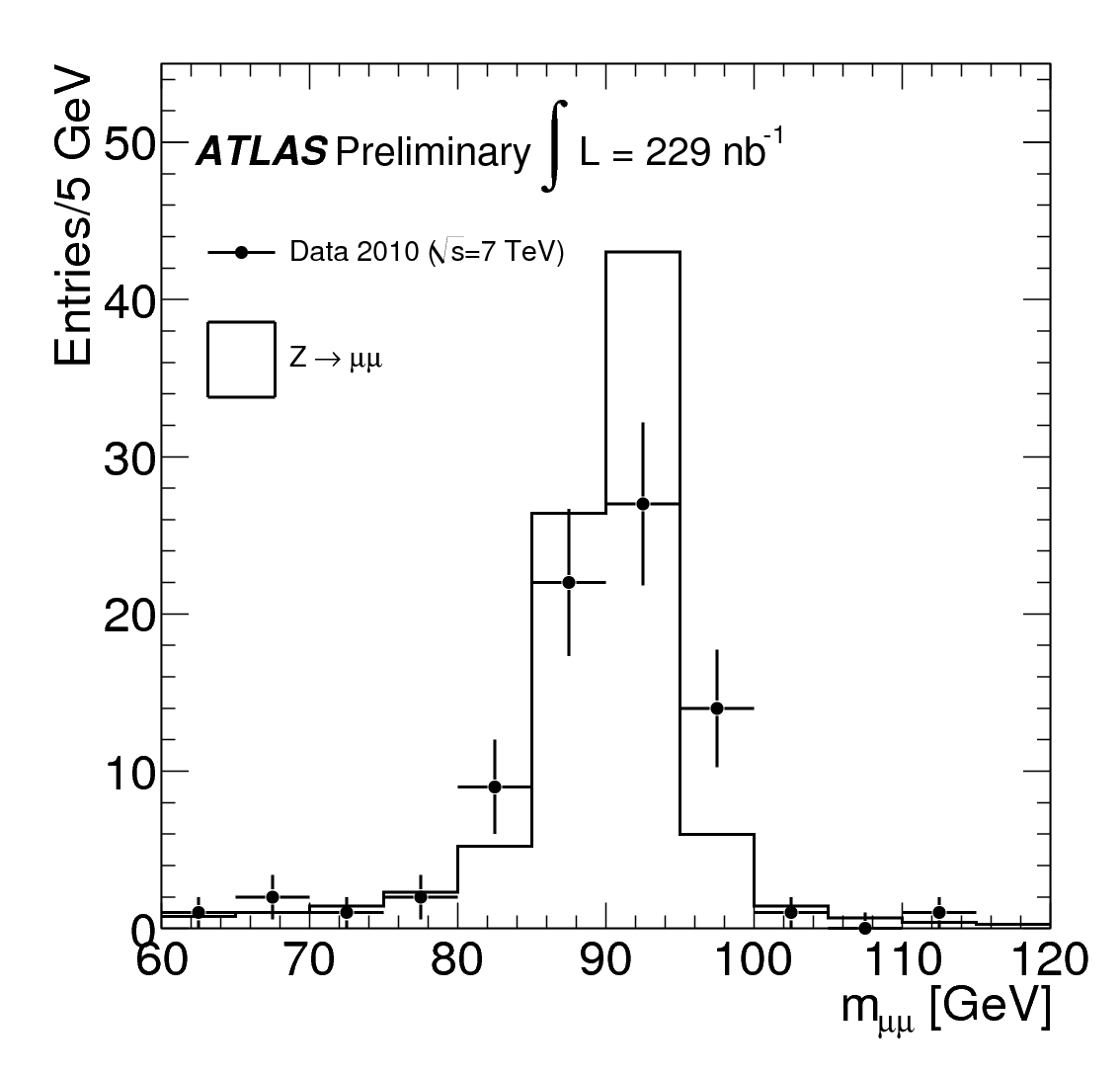}
} \\
\caption{Invariant mass $m_{\mathrm{ll}}$ of Z candidates for $ee$ (a)
  and $\mu\mu$ (b). The background is suppressed because of
  its smallness. \label{fig:mll}}
\end{figure}

The $W$ analysis requires that the transverse mass $m_{\mathrm{T}}$ of the
candidate $l$ - $E_{\mathrm{T}}^{\mathrm{Miss}}$ system, plotted in
Fig.~\ref{fig:mtw}, be $m_{\mathrm{T}} \ge 40\,\mathrm{GeV}$. There are $46$ ($72$)
$\mathrm{W}\to e \nu$ ($\mu\nu$) candidates. Electroweak
(EW) backgrounds for $W$ are obtained from Monte Carlo
predictions. In the electron channel, the EW background is
$1.5\pm0.0\,\mathrm{(stat)}\pm0.1\,\mathrm{(syst)}$, coming chiefly
from leptonic $\mathrm{W}\to\tau\nu$. The muon channel
background of $4.4\pm0.0\,\mathrm{(stat)}\pm0.3\,\mathrm{(syst)}$, is
dominated by $\mathrm{W}\to\tau\nu$ and $\mathrm{Z}\to\mu\mu$. The
QCD background in $\mathrm{W}\to e\nu$ comes from hadrons
misidentification, as well as electrons from conversion and heavy
quark decay. It is estimated with a binned
likelihood fit to data of Monte Carlo templates of calorimetric isolation, at
$1.1\pm0.2\,\mathrm{(stat)}\pm0.4\,\mathrm{(syst)}$ events. The muon
channel QCD background, mainly from heavy quark decay, is estimated by extrapolating from control
regions in the $E_T^{\mathrm{Miss}}$ vs. isolation plane, at
$0.9\pm0.3\,\mathrm{(stat)}\pm0.6\,\mathrm{(syst)}$. $Z$ boson candidates are formed using pairs of oppositely charged
leptons, with the invariant mass of the lepton pair required to be
between $66\,\mathrm{GeV}$ and $116\,\mathrm{GeV}$. 46 (79) candidates are observed in the
$\mathrm{Z}\to e^+ e^-$ ($\mu^+ \mu^-$) channel, whose
peak is shown in Fig.~\ref{fig:mll}. The
background for $Z \to e^+ e^-$ is $0.49\pm
0.07\,\mathrm{(stat)}\pm0.05\,\mathrm{(syst)}$. The background in the
muon channel is $0.17\pm0.01\,\mathrm{(stat)}\pm0.01\pm\mathrm{(syst)}$.

\section{CROSS-SECTION AND CONCLUSION}
For $N$ observed events and background estimate of $N^{\mathrm{bkg}}$
in a luminosity of $L^{\mathrm{int}}$, we have:
\begin{equation}
\sigma_{\mathrm{\mathrm{W}
    (\mathrm{Z})}} \times \mathrm{BR}_{\mathrm{W}\to l\nu
  (\mathrm{Z} \to ll)} = \frac{N - N^{\mathrm{bkg}}_{\mathrm{W}
    (\mathrm{Z})}}{A_{\mathrm{W}
    (\mathrm{Z})} C_{\mathrm{W}
    (\mathrm{Z})} L^{\mathrm{int}}_{\mathrm{W}
    (\mathrm{Z})}}
\end{equation}
Here, $A$ is the geometrical acceptance at generator
level, while $C$ is the ratio between the number of selected
reconstructed events and the number of generated events in the
fiducial region. The systematics on $A_{\mathrm{W}}$ and
$A_{\mathrm{Z}}$ are estimated at $3\%$, from using different PDF sets
and comparing MC@NLO and PYTHIA values. The uncertainty on $C$ in the
electron channel is $8\%$
and $14\%$ for $\mathrm{W}$ and $\mathrm{Z}$, driven by the lepton identification. The muon systematic comes from
the trigger, muon reconstruction and resolution, and from missing energy for the $\mathrm{W}$. It is $7\%$ in both the
$\mathrm{W}$ and $\mathrm{Z}$ analyses.

\begin{figure}[h!]
\centering
\subfigure[$\mathrm{W}\to l \nu$] {
\label{fig:wxsec}
\includegraphics[width=.4\textwidth]{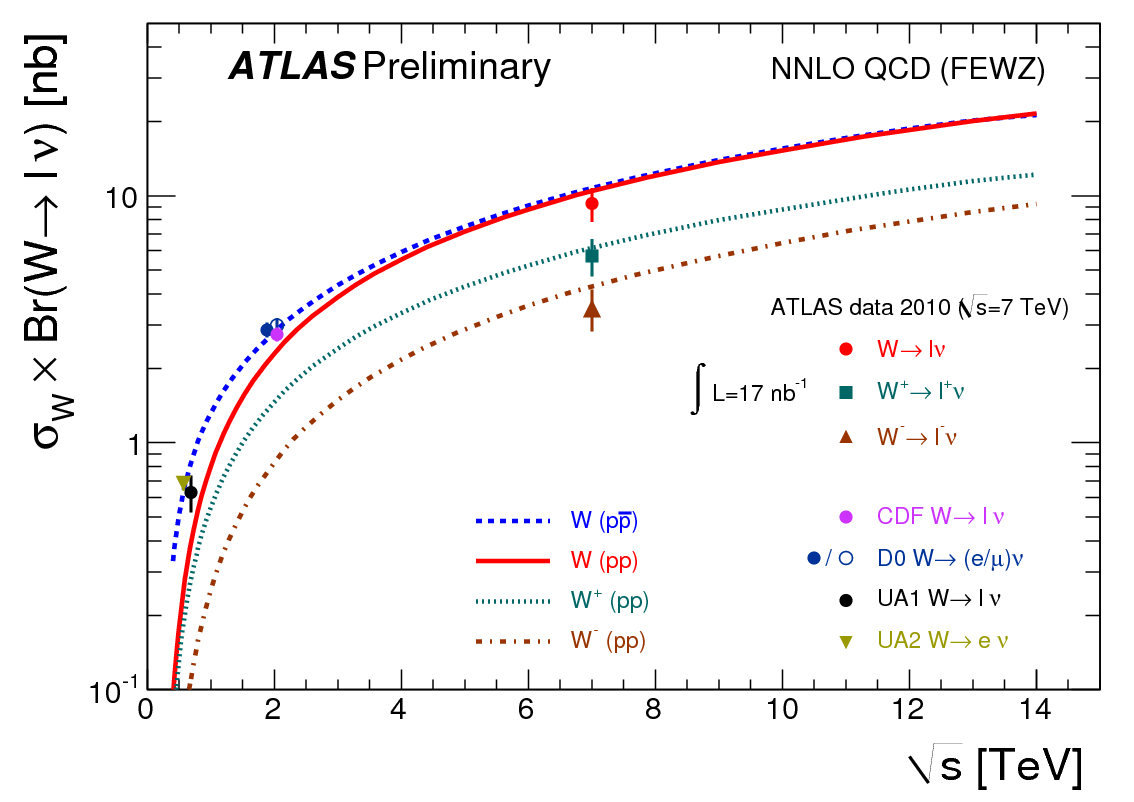}
}
\subfigure[$\mathrm{Z}\to l^+l^-$]{
\label{fig:zxsec}
\includegraphics[width=.4\textwidth]{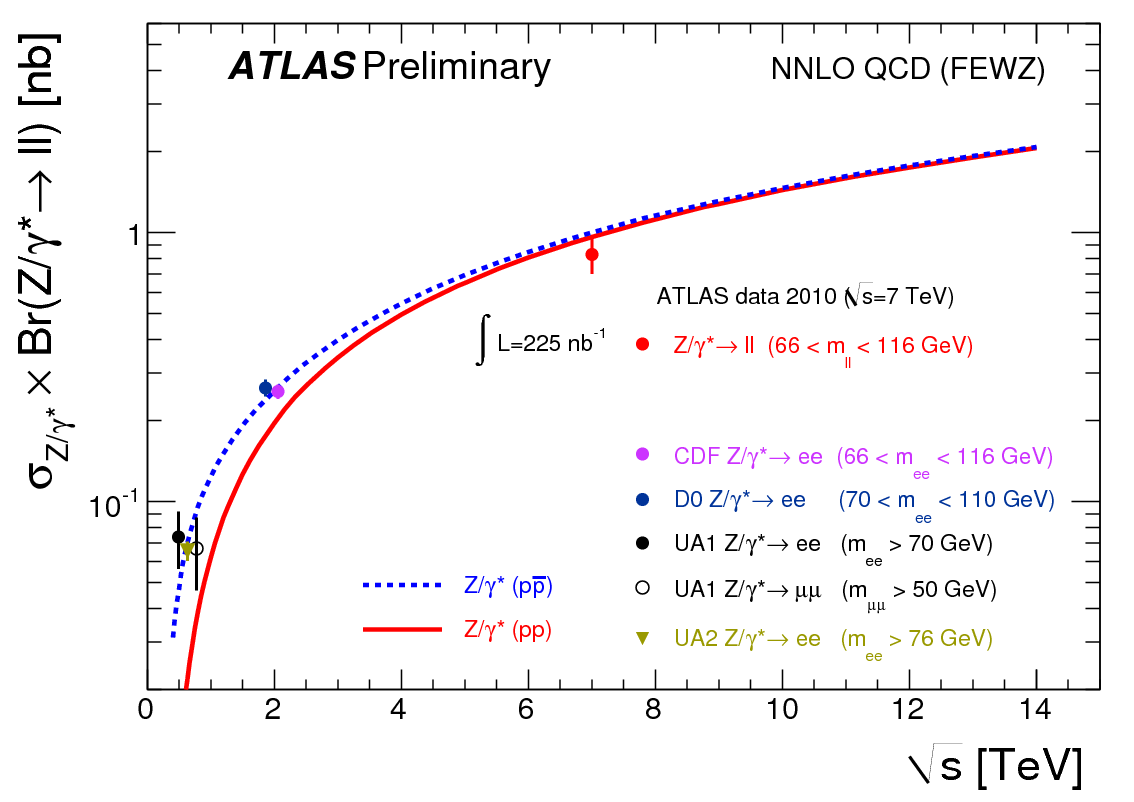}
} \\
\caption{The measured values $\sigma_{\mathrm{W}} \times\,\mathrm{BR }
  \mathrm{W} \to l \nu$ (a) and $\sigma_{\mathrm{Z}}
  \times\,\mathrm{BR }
  \mathrm{Z/\gamma^*} \to l^+l^-$ (b), compared to NNLO
  QCD calculations and to previous measurements. The predictions are
  shown for both $pp$ and $p\bar{p}$ colliders as a function of $\sqrt{s}$.
 \label{fig:vbxsec}}
\end{figure}

The cross-section times leptonic branching fraction for $\mathrm{W}$ is $\left[8.5\pm1.3\,\mathrm{(stat)}\pm0.7\,\mathrm{(syst)}\pm0.9\,\mathrm{(lumi)}\right]\,\mathrm{nb}$
in the electronic channel and $\left[10.3\pm1.3\,\mathrm{(stat)}\pm0.8\,\mathrm{(syst)}\pm1.1\,\mathrm{(lumi)}\right]\,\mathrm{nb}$
for the muonic channel. The combined value is
$\left[9.3\pm0.9\,\mathrm{(stat)}\pm0.6\,\mathrm{(syst)}\pm1.0\,\mathrm{(lumi)}\right]\,\mathrm{nb}$. The
$\mathrm{Z}\to e^+e^-$ and $\mathrm{Z}\to\mu^+\mu^-$ results are
respectively
$\left[0.72\pm0.11\,\mathrm{(stat)}\pm0.10\,\mathrm{(syst)}\pm0.08\,\mathrm{(lumi)}\right]\,\mathrm{nb}$
and
$\left[0.89\pm0.10\,\mathrm{(stat)}\pm0.07\,\mathrm{(syst)}\pm0.10\,\mathrm{(lumi)}\right]\,\mathrm{nb}$,
for a combined result of
$\left[0.83\pm0.07\,\mathrm{(stat)}\pm0.06\,\mathrm{(syst)}\pm0.09\,\mathrm{(lumi)}\right]\,\mathrm{nb}$. These
results are placed in historical context in Fig.~\ref{fig:vbxsec}, and
in agreement with theoretical expectations of $\sigma_{W\to l\nu} =
10.46\,\mathrm{nb}$, and $\sigma_{Z/\gamma^{*}\to ll} =
0.99\,\mathrm{nb}$, computed at NNLO with 4\% uncertainty using FEWZ
\cite{fewz} and MSTW2008 \cite{mstw08}.

\end{document}